# Effect of Ti doping on the electrical transport and magnetic properties of layered compound $Na_{0.8}CoO_2$

W. Y. Zhang, Y. G. Zhao,[a)] Z. P. Guo, P. T. Qiao, L. Cui, L. B. Luo, and X. P. Zhang

*Department of Physics, Tsinghua University, Beijing 100084, China*

H. C. Yu, Y. G. Shi, S. Y. Zhang, T. Y. Zhao, and J. Q. Li

*Beijing National Laboratory for Condensed Matter Physics, Institute of Physics, Chinese Academy of Sciences, Beijing 100080, China*

**Abstract**

Effect of Ti doping on the electrical transport and magnetic properties of layered $Na_{0.8}Co_{1-x}Ti_xO_2$ compounds has been investigated. The lattice parameters *a* and *c* increase with x. A minor amount of Ti doping results in a metal-insulator transition at low temperatures. For samples with x > 0.03, the variable-range hopping process dominates the transport behavior above a certain temperature. The temperature dependence of magnetization of all the samples is found to obey the Curie-Weiss law. The mechanism of the doping effect is discussed.

# Introduction

Hexagonal $Na_xCoO_2$ compounds have stimulated intense investigations due to the existence of their unique crystal structure, rich physical properties and potential thermoelectric applications [1-3]. $Na_xCoO_2$ has a two-dimensional layered structure where Na ions and $CoO_2$ layers are alternately stacked. The Co ions form the two-dimensional triangular lattice, which may be the realization of Anderson's original triangular lattice RVB system. Charge-ordering is found to exist in $Na_{0.5}CoO_2$ compound. The room-temperature thermopower coefficient of $Na_xCoO_2$ compounds is almost ten times higher than that of typical metals, however, their resistivity is only about 200 μΩmm.

$Na_xCoO_2$ compounds have a very large electronic specific-heat coefficient γ [4]. The Hall coefficient has an opposite sign to the thermopower and is strongly temperature dependent [5]. These phenomena indicate that $Na_xCoO_2$ compounds belong to the strongly correlated system. In this kind of systems, interaction among spin, charge, and orbital plays a major role in producing the peculiar physical properties of these compounds, as in the case of high-$T_c$ cuprates. The tiny change of composition of the $Na_xCoO_2$ compounds easily affects spin-charge-orbital interaction and results in a dramatic change of their physical properties. For example, a small amount of Ca doping effectively decreases the carrier concentration and significantly increases the thermopower and resistivity of $Na_{1.1-x}Ca_xCo_2O_2$ [6]. 3% Mn doping for $NaCo_{1-x}Mn_xO_2$ induces a metal-insulator transition possibly arising from the magnetic disorder effect [7]. The above results show that the physical properties of $Na_xCoO_2$ compounds are very sensitive to the doping effect. In addition, The LDA+U study by Zhang et al indicates that the electronic and structure properties of $Na_xCoO_2$ are also dependent on the element doping [8]. In order to get more insight into the mechanism of doping effect of $Na_xCoO_2$, it is interesting to explore the doping effect of other elements on the physical properties of $Na_xCoO_2$.

In this article, the effect of Ti doping on the structural, magnetic and electrical transport properties of $Na_xCoO_2$ compounds has been investigated. It shows that the electrical transport property of $Na_xCoO_2$ is very sensitive to Ti doping. Metal-insulator transition occurs at low temperatures for 3% doped sample. The electrical transport of the Ti-doped sample can be described by the 3D variable-range hopping (VRH) mechanism. The doping effect was discussed by considering some possible mechanisms.

**Experimental**

Polycrystalline samples with nominal compositions $Na_{0.8}Co_{1-x}Ti_xO_2$ (x = 0, 0.03, 0.05, 0.07, 0.1, 0.2, 0.3, 0.4) were prepared by the solid-state reaction. A stoichiometric amount of $Na_2CO_3$, $Co_3O_4$, and $TiO_2$ was mixed and sintered at 750~860 $^0$C for 10 h in air. The product was finely ground, pressed into pellets, and sintered at 800-900 $^0$C for 8 h in air. Since Na tends to evaporate during calcinations [9], the undoped sample with the starting composition of $Na_{0.8}CoO_2$ is expected to change to about $Na_\gamma CoO_2$ with 0.70<$\gamma$<0.75.

X-ray diffraction was performed using a Rigaku D/max-RB x-ray diffractometer with Cu $K_\alpha$ radiation as the x-ray source in the θ-2θ scan mode. The electrical resistivity $\rho(T)$ was measured using the four probe method within the temperature range of 5~300 K. Indium was used for the electrical contact. The dc magnetization was measured using a superconducting quantum interference device (SQUID) magnetometer within the temperature range of 5~300 K in a 500 Oe magnetic field.

**Results and discussions**

Fig.1 shows the XRD patterns for $Na_{0.8}Co_{1-x}Ti_xO_2$. Because the peak at 16.5 ° is very strong, the figure was shown with two separate figures. No impurity phase is detected within the experimental resolution, which is about 3%. All the diffraction peaks can be indexed according to the γ phase [10], so Ti is fully substituted for Co. It is concluded that the solid solubility limit of Ti in $Na_{0.8}CoO_2$ is larger than 0.4, which is much larger than the solid solubility limit of Cu in $Na_{0.5}CoO_2$ [11]. Single phased $Na_{0.7}Co_{1-x}Ti_xO_2$ samples were also obtained for x≤0.1 [12]. The values of the lattice constant *a* and *c* are presented in Fig.2. With the increase of x, the lattice parameters *a* and *c* increase. This is expected because the ions radius of Ti is bigger than that of Co. Similar results are observed in $NaCo_{1-x}Mn_xO_2$ [7].

In order to study the influence of the Ti doping on the valence states of Co in the $Na_{0.8}Co_{1-x}Ti_xO_2$ materials, we have performed EELS analyses on x = 0 and x = 0.2 samples. Figure 3(a) shows the EELS spectra of $Na_\gamma CoO_2$ and $Na_\gamma Co_{0.8}Ti_{0.2}O_2$ materials taken from an area of about 100 nm in diameter. In these spectra, the typical peaks, i.e. the collective plasmon peaks as well as core edges for Co, Ti, and O elements, are displayed. Figure 3(b) show a typical spectra for the Co $L_2$ and $L_3$ peaks obtained from x=0 and 0.2 samples. According to the value of $L_3/L_2$, it is determined that the Co valence is increased from 3.3 for x = 0 to 3.5 for x = 0.2. So we conclude that Ti may replace the $Co^{3+}$ ions.

Fig.4 presents the temperature dependence of resistivity for $Na_{0.8}Co_{1-x}Ti_xO_2$. For the sample with x = 0.03, the resistivity decreases linearly with the decrease of temperature from 300 K to 25 K, showing the metallic behavior. Further decrease of temperature results in the abrupt increase of resistivity, which is a metal-insulator transition. Similar temperature dependence of the resistivity is also observed in $Ca_3Co_4O_9$ compounds [13]. For the sample with x > 0.03, the resistivity increases with the decrease of temperature, showing the semiconductor-like behavior. It is no doubt that a small amount of Ti doping leads to a metal-insulator transition, which is similar to that in $NaCo_{1-x}Mn_xO_2$ [7]. It can be seen from the inset that the resistivity at the temperature of 15 K increases with the increase of x and an abrupt increase occurs at about x=0.07.

The semiconductor-like behavior of resistivity is generally characterized by three models, i.e., a band-gap model [14], a nearest-neighbor hopping model [15], and VRH model [16]. We fitted the experimental data for the sample with x = 0.05 by using the above three models, respectively. The three-dimensional (3D) VRH model gives a satisfactory fit in a wide temperature range. This shows that the VRH process dominates the conduction behavior. We also employ the 3D VRH model to fit the transport behavior of other samples and the fitting curves are shown in Fig.5. It is evident that the curves of all the samples are linear within a wide temperature range. This indicates that the transport behavior of the Ti-doped samples is mainly controlled by the 3D VRH mechanism. For 3D VRH mechanism, $\rho(T)= \rho_0 \exp(T_0/T)^{1/4}$, where $T_0=\beta/[k_B N(0)d^3]$, $N(0)$ is the density of states at the Fermi level, $d$ is the localization radius of states near the Fermi level, $k_B$ is Boltzmann's constant, and $\beta$ is a numerical coefficient. Similar to the doing dependence of resistivity, the doping dependence of $T_0$ also shows an abrupt increase at x=0.07 as shown in the inset of Fig.5. "Incoherence-coherence" transition has been found in the electrical transport in $(Bi_{0.5}Pb_{0.5})_2Ba_3Co_2O_y$ and $NaCo_2O_4$ at 200 K and 180 K, respectively [3, 17]. This "incoherence-coherence" transition corresponds to a crossover from two to three dimensional transport behavior. Because the value of lattice parameter $c$ for $Na_{0.8}Co_{1-x}Ti_xO_2$ becomes larger with x increasing, the corresponding temperature for dimensional crossover from 2D to 3D will become low. So it is not incomprehensible that the start temperature of the 3D VRH process decreases with the increase of the Ti content.

Fig.7 shows the result of dc magnetization for the $Na_{0.8}Co_{1-x}Ti_xO_2$ samples. All the samples exhibit Curie-Weiss behavior, indicating the existence of localized moment. Using the Curie-Weiss law $\chi=\chi_0+C/(T-\theta)$ to fit the experimental data, where $\theta$ is the paramagnetic Curie-Weiss temperature, $\chi_0$ is a sum of temperature-independent terms, and $C = N\mu_{eff}^2/3k_B$ ($\mu_{eff}$ is the effective moment of the magnetic ions, $k_B$ is the Boltzman constant, and N is the number of magnetic ions per unit volume), we can get the effective magnetic moment of the system. For x=0, 0.10, 0.20 samples, the effective magnetic moment per molecule is 0.723 $\mu_B$, 0.931 $\mu_B$ and 1.129 $\mu_B$, respectively. So the effective magnetic moments of the samples increase with Ti doping. For x=0, 0.10, 0.20 samples, the Weiss temperature ($\theta$) is -123 K -282 K, -309 K, respectively. This implies that the magnetic interaction increases with Ti doping. The changes of the effective magnetic moment and the Weiss temperature with Ti doping may be attributed to the substitution of nonmagnetic $Co^{3+}$ ions by magnetic Ti ions. This is consistent with the EELS data.

As shown by EELS data, Ti doping increases the Co valence. The increase of the Co valence is due to the substitution of $Co^{3+}$ ions by Ti ions, so the number of $Co^{4+}$ may be unchanged. Since the hole carriers are contributed by $Co^{4+}$ ions, the carrier density in the samples may be fixed. So other factors should be considered for the electronic transport in the Ti doped samples. It has been shown that the distance between Co in the $CoO_2$ layers is very important in determining the electrical transport property of $NaMO_2$ (M=Cr, Mn, Fe, Co, Ni) and a critical M-M distance $R_c$ can be used to judge whether the sample is a metal or an insulator [18]. According to the model proposed by Goodenough, $R_c$=[3.20-0.05m-0.03(Z-$Z_{Ti}$)-0.04s(s+1)]Å, where m is the valency of $M^{m+}$, Z is the atomic number of the transition metal, s is the effective spin. Molenda [19] used this formula to analyse the electrical transport property of $A_xMO_2$ (A=Li, Na; M=Cr, Mn, Fe, Co, Ni) and found that the comparison of the real metal-metal (M-M) distance (determined by experimental) with the critical M-M distance given by Goodenough's model can predict the property of the materials. For example, $R_c$ is in the range of 2.734-2.727 Å for $Na_xMnO_2$ with 0.7≤ x ≤0.85, and the real M-M distance is in the range of 2.87-2.88 Å, which is much larger than $R_c$. So the model predicts that $Na_xMnO_2$ is a semiconductor which is consistent with the electrical transport property of $Na_xMnO_2$. While for $Na_{0.7}CoO_2$, $R_c$ is about 2.822 Å, and the real M-M distance is in the range of 2.815 Å, which is very close to $R_c$. So $Na_{0.7}CoO_2$ is a metal, which is also consistent with experimental observation. From above discussion, it can be deduced that Mn doping tends to decrease $R_c$ and increase the real M-M distance. This will drive the system to become a semiconductor and this is supported by our previous work [7]. Ti doping makes the lattice parameter *a* of $Na_{0.8}Co_{1-x}Ti_xO_2$ increase which leads to the increase of the distance between Co in the $CoO_2$ layers. So Ti doping is detrimental to the metallic conductivity of $Na_{0.8}Co_{1-x}Ti_xO_2$ in terms of Co-Co distance. We believe that for lightly doped samples, the disorder caused by doping plays an important role in determining the electrical transport property of the samples. For example, Ti doping disturbs the magnetic interaction in the $CoO_2$ layers, and magnetic disorder is also detrimental to the carrier transport. This disorder may cause the localization of carriers at low temperatures. The doping of magnetic Ti ions in a metallic system may also lead to the Kondo effect, which shows a resistivity minimum at low temperatures. For samples with large doping, the arguments based on Goodenough's model may play a major role, i.e. doping leads to the increase of M-M distance and exceeds $R_c$, resulting in a semiconductor behavior.

**Conclusions**

Layered $Na_{0.8}Co_{1-x}Ti_xO_2$ samples have been prepared. The solid solution limit of Ti is more than 0.4. The Co valence is demonstrated to increase with the increase of Ti content indicating that Ti may mainly substitute the $Co^{3+}$ ions. Ti doping results in a metal-insulator transition, which may arise from the magnetic disorder effect induced by Ti doping. The transport behavior of the samples with Ti doping is mainly controlled by the 3D variable-range hopping mechanism. The temperature dependence of magnetization of all the samples obeys the Curie-Weiss law and the effective moment is increased due to the introduction of Ti. This work demonstrates that layered $Na_{0.8}CoO_2$ compound is very sensitive to magnetic ions doping. This implies that magnetic interaction plays an important role in this layered compound.

## Acknowledgements

This work was supported by NSFC (Nos.50272031 and 50425205), the Excellent Young Teacher Program of MOE, P.R.C, National 973 project (No. 2002CB613505) and Specialized Research Fund for the Doctoral Program of Higher Education (No. 2003 0003088).

**Figure captions**

Fig.1 XRD patterns for $Na_{0.8}Co_{1-x}Ti_xO_2$.

Fig.2 The change of the lattice constant *a* and *c* vs x.

Fig.3 (a) EELS spectra of $Na_\gamma CoO_2$ and $Na_\gamma Co_{0.8}Ti_{0.2}O_2$ materials, (b) Typical spectra for the Co $L_2$ and $L_3$ peaks obtained from the samples with x=0 and 0.2.

Fig.4 Temperature dependence of resistivity for $Na_{0.8}Co_{1-x}Ti_xO_2$.

Fig.5 R-T Curves fitted by 3D VRH model for the samples with x = 0.05, 0.07, 0.1, respectively. The inset is the doping dependence of $T_0$.

Fig.6 Temperature dependence of magnetization for the $Na_{0.8}Co_{1-x}Ti_xO_2$ samples.

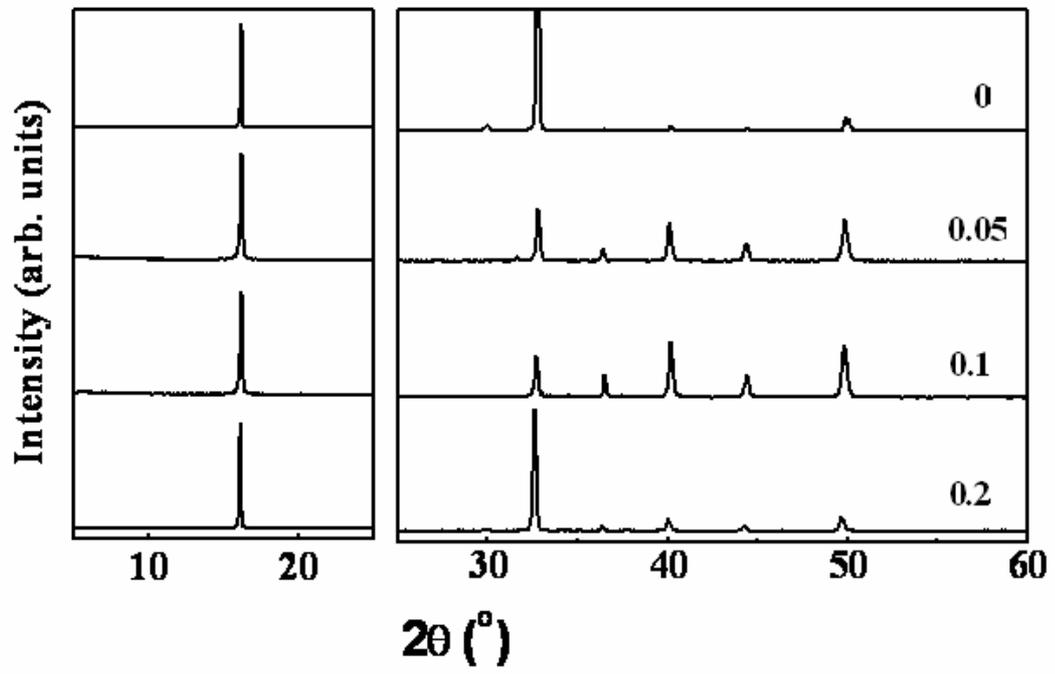

**Fig.1**

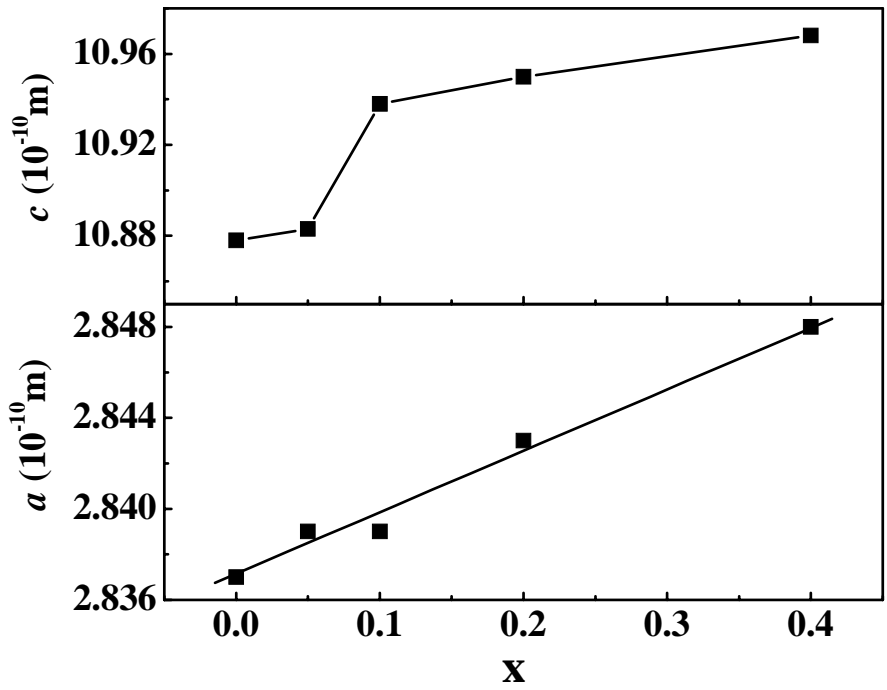

**Fig.2**

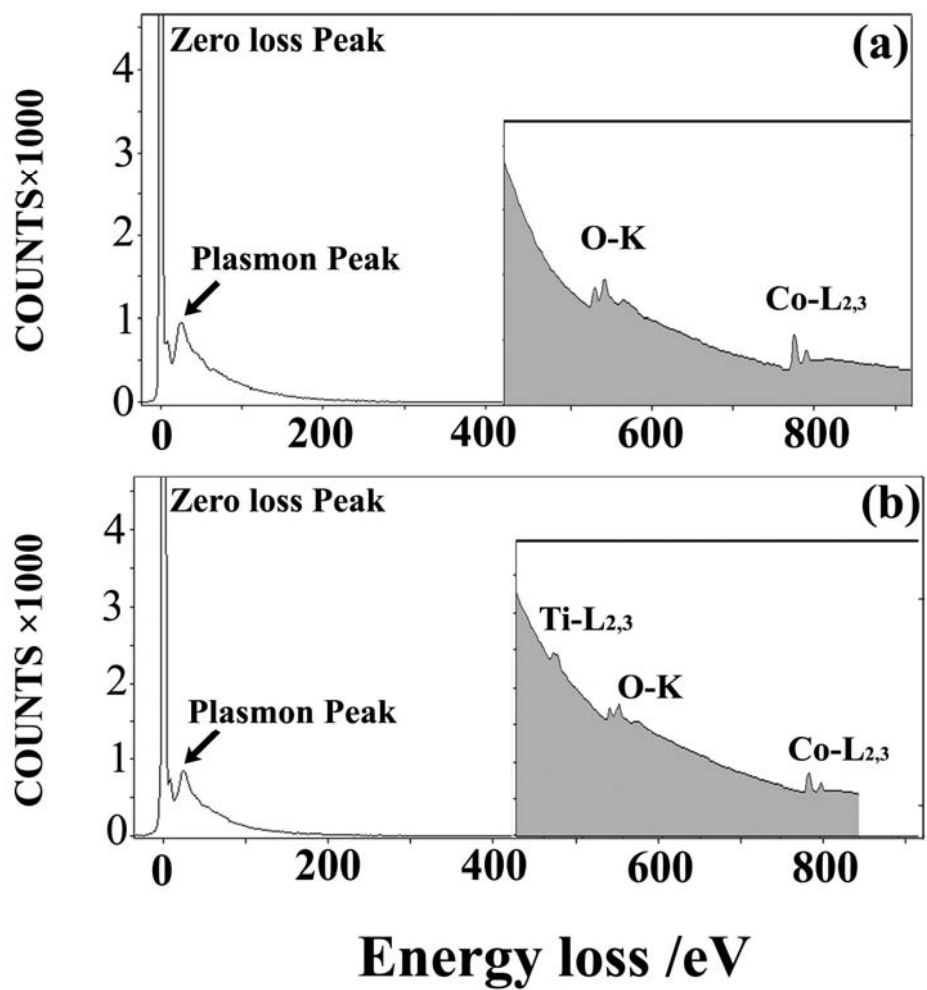

**Fig.3 (a)**

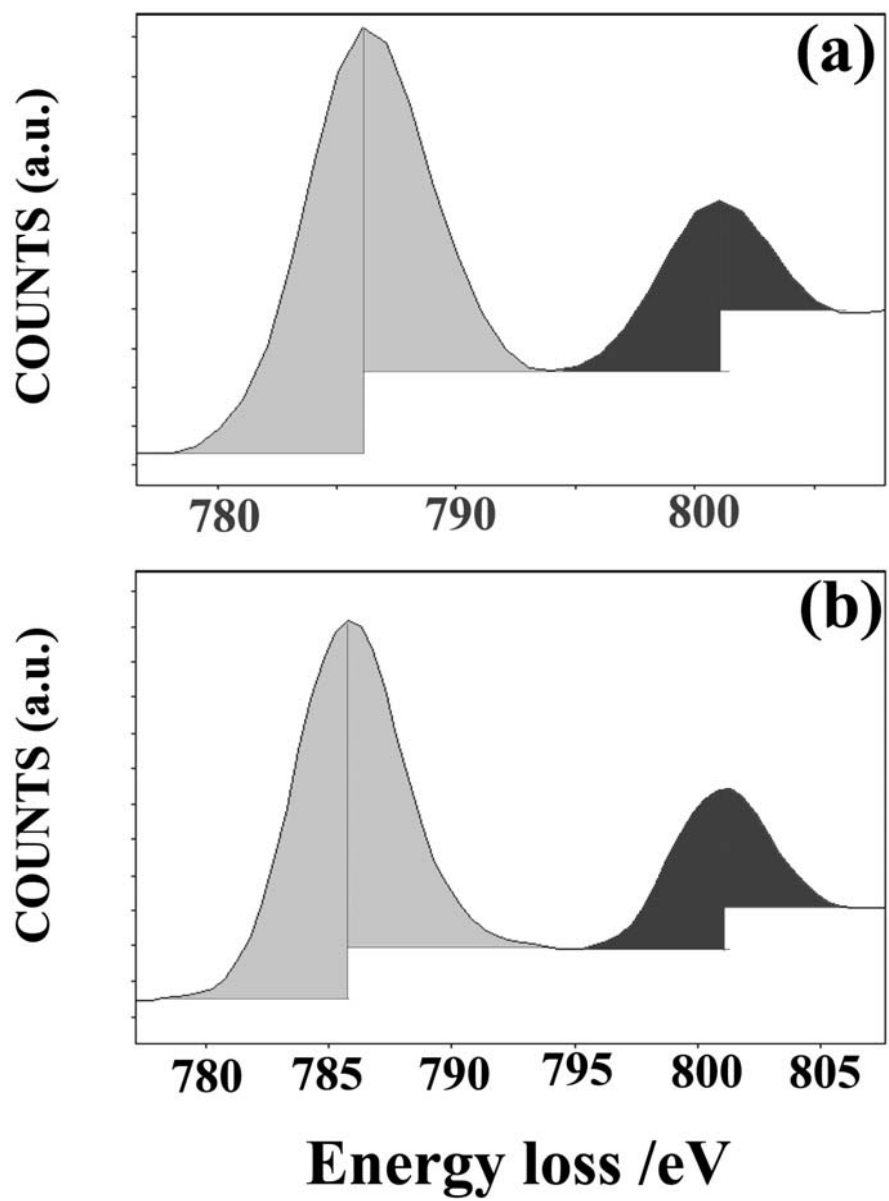

**Fig.3 (b)**

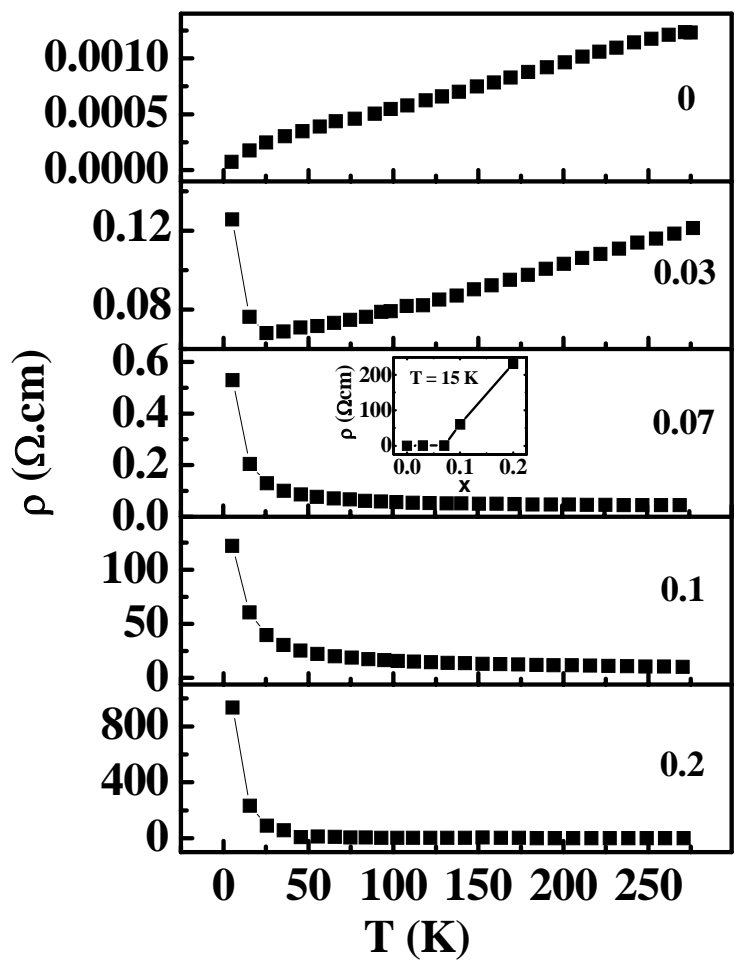

**Fig.4**

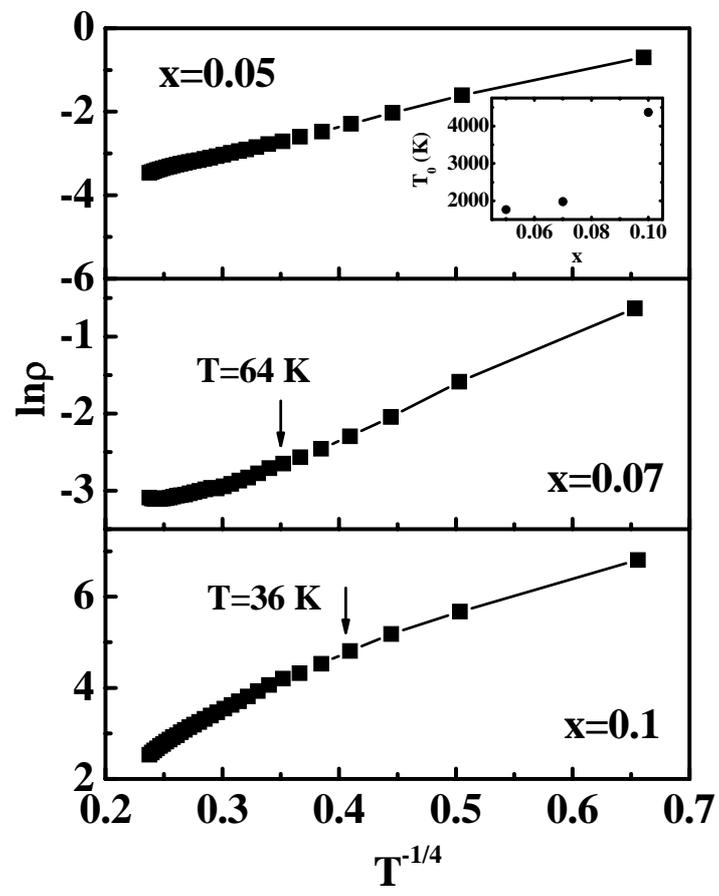

**Fig.5**

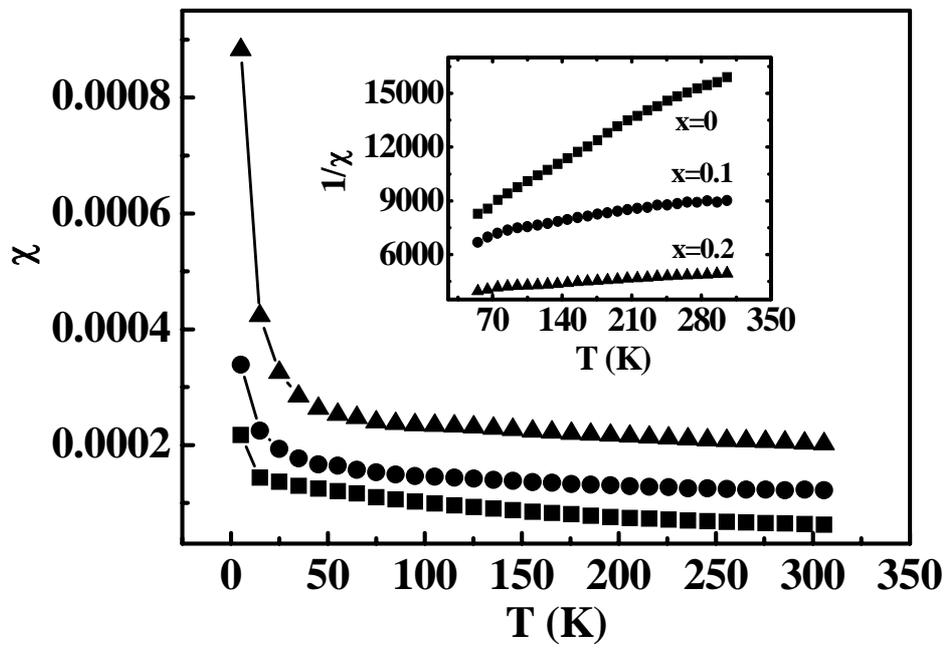

**Fig.6**